\documentclass[eat,twocolumn]{jmlr}
   
\usepackage{longtable}

\usepackage{booktabs}
\usepackage[load-configurations=version-1]{siunitx} 

\theorembodyfont{\upshape}
\theoremheaderfont{\scshape}
\theorempostheader{:}
\theoremsep{\newline}

\jmlrvolume{1}
\firstpageno{1}

\jmlryear{2022}
\jmlrworkshop{Machine Learning for Health (ML4H) 2022}

\newcommand{\err}[1]{\color {gray}\scriptsize $\pm #1$}

\title[Short Title]{
Improved 
proteasomal cleavage 
prediction 
with
positive-unlabeled learning
}

\author{
 \Name{Emilio Dorigatti} \Email{edo@stat.uni-muenchen.de} \\
 \addr Ludwig Maximilian Universit\"at M\"unchen,
 \addr Helmholtz Zentrum M\"unchen
 \AND
 \Name{Bernd Bischl} \Email{bernd.bischl@stat.uni-muenchen.de} \\
 \addr Ludwig Maximilian Universit\"at M\"unchen,
 \addr Munich Center for Machine Learning
 \AND
 \Name{Benjamin Schubert} \Email{benjamin.schubert@helmholtz-muenchen.de} \\
 \addr Helmholtz Zentrum M\"unchen,
 \addr Technical University of Munich
}

\begin{document}
\maketitle
\begin{abstract}
Accurate \emph{in silico} modeling of the antigen processing pathway is crucial to enable personalized epitope vaccine design for cancer.
An important step of such pathway is the degradation of the vaccine into smaller peptides by the proteasome, some of which are going to be presented to T cells by the MHC complex.
While predicting MHC-peptide presentation has received a lot of attention recently, proteasomal cleavage prediction remains a relatively unexplored area in light of recent advances in high-throughput mass spectrometry-based MHC ligandomics.
Moreover, as such experimental techniques do not allow to identify regions that cannot be cleaved, the latest predictors generate decoy negative samples and treat them as true negatives when training, even though some of them could actually be positives.
In this work, we thus present a new predictor trained with an expanded dataset and the solid theoretical underpinning of positive-unlabeled learning, achieving a new state-of-the-art in proteasomal cleavage prediction.
The improved predictive capabilities will in turn enable more precise vaccine development improving the efficacy of epitope-based vaccines.  
Pretrained models are available on GitHub.\footnote{ \url{https://github.com/SchubertLab/proteasomal-cleavage-puupl}}
\end{abstract}


\begin{figure*}
    \centering
    \includegraphics[width=0.99\textwidth]{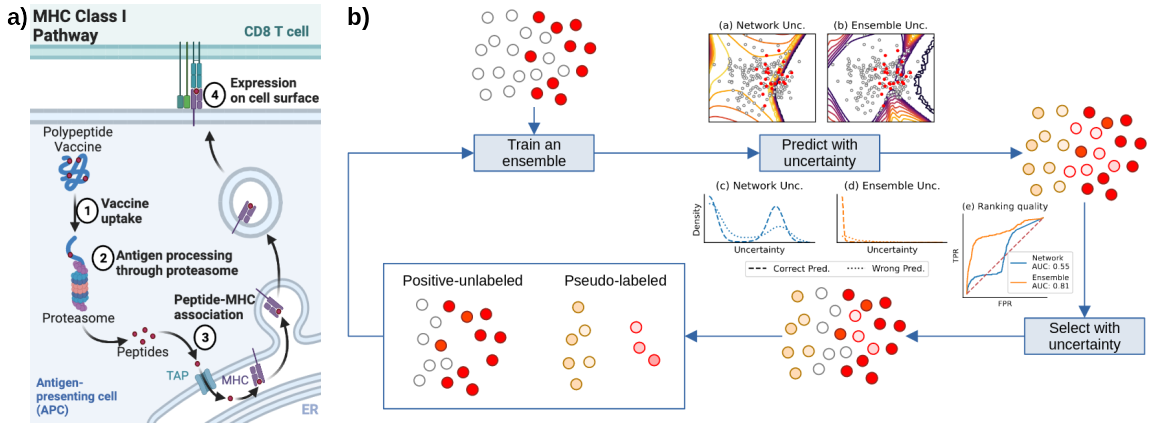}
    \caption{
    \small
    a) The antigen-processing pathway: after uptake of the vaccine (1), the proteasome cleaves it into smaller pieces (2), some of which are loaded onto a MHC molecule (3) and subsequently presented on the surface of the cell (4).
    b) The iterative PUUPL framework for positive-unlabeled learning. The confident predictions of an ensemble are mostly correct, and are thus used as pseudo-labels for for the next training iterations, continuously improving the model's performance.
    }
    \label{fig:antiprocpath}
\end{figure*}

\section{Introduction}
Epitope-based cancer vaccines (EV) target neoepitopes that arise in tumors as a result of somatic mutations.
Such vaccines are designed by identifying a subset of neoepitopes with maximal immunogenicity that can be joined by short linker sequences designed \emph{ad hoc} to facilitate processing of the resulting polypeptide by the antigen processing pathway of the human immune system~\citep{jessev,genev,toussaint_universal_2011}
A core step of such pathway (Figure~\ref{fig:antiprocpath}a) is the digestion of the polypeptide by the proteasome, a protein complex which degrades old or damaged proteins into fragments, some of which are then loaded onto the Major Histocompatibility Complex (MHC) and transported to the surface of the cell for presentation to T cells~\citep{antigen_pathway}.
To ensure effectiveness at low doses, the design of EVs has to balance immunogenicity of the chosen neoepitopes with the likelihood that such neoepitopes will be correctly recovered after proteasomal cleavage.
In order to increase efficiency and reduce production costs of EVs it is thus of paramount importance to model proteasomal cleavage in order to predict the resulting fragments, and design the vaccine so that these fragments correspond to the desired neoepitopes~\citep{jessev}.

Although it is possible to conduct \emph{in vitro} experiments to collect data on proteasomal cleavage, such process is time-consuming, expensive, and low-throughput, thus only scarce data has been collected so far.
At the same time, the development of high-throughput pipelines for mass spectrometry-based MHC ligandomics produces vast amount of data on peptide presentation by the MHC~\citep{massspec,iedb}.
Crucially, all peptides detected in such a way must have resulted from proteasomal cleavage, providing indirect information on that process.
However, peptides with low MHC binding affinity are often not detected, and information about missed cleavage sites is also unavailable, thus complicating the development of cleavage predictors from this kind of data.
Positive-unlabeled learning (PUL) is a branch of semi-supervised learning which is concerned with learning from datasets missing negative labeled examples, but having only positives and unlabeled data~\citep{bekker2020}.
PUL is thus a natural fit for training proteasomal cleavage predictors based on MHC ligandomics data.

In this work, we combine the newly available MHC ligandomics datasets with our recent PUL framework PUUPL~\citep{puupl} to develop an updated proteasomal cleavage predictor with considerably higher performance than currently available.

\section{Related work}

\paragraph{Epitope vaccine design}
was first approached by optimization frameworks that approached epitope selection and assembly separately~\citep{lundegaard_popcover_2010, toussaint_mathematical_2008},
first using pre-determined spacer sequences \citep{veldersDefinedFlankingSpacers2001} then spacers specifically optimized for each epitope pair to maximize their cleavage likelihood\citep{schubert_designing_2016}.
The latest EV design frameworks approach both stages concurrently as a multi-objective optimization where vaccine immunogenicity is optimized jointly with cleavage likelihood~\citep{genev,jessev}.

\paragraph{Proteasomal cleavage prediction}
was approached with neural network models~\citep{netchop_first,Kuttler2000} as well as linear predictors while jointly modeling all steps of the antigen processing pathway~\citep{pcm}.
Subsequently, predictors were developed for other proteases~\citep{mcpred,deepdigest}, however they are not applicable to immune proteasomal cleavage prediction as the underlying biology is different.

NetChop~\citep{Nielsen_2005} is a multi-layer-perceptron (MLP)-based cleavage predictor trained in a traditionally supervised setting on a similar, albeit much smaller, dataset consisting of two \textit{in vitro} proteasomally digested proteins as well as another dataset consisting of MHC-I ligands, and requiring a window of 17 amino acids around the potential cleavage site.
NetCleave~\citep{netcleave}, also based on MLPs was presented recently and provides prediction for specific MHC classes or alleles while requiring a window of only seven amino acids.

\paragraph{Positive-unlabeled learning}
was first introduced as a variant of binary classification \citep{Liu2003} and approached with a variety of methods~\citep{bekker2020}.
An unbiased risk estimator supported by solid theoretical foundations was introduced by~\citet{plessis2014} and later improved by~\citet{Kiryo2017} for deep learning methods.
In this work, we use PUUPL~\citep{puupl}, a recently proposed general framework for PUL that is suitable for sequence data, has competitive performance on imbalanced datasets and provides naturally well-calibrated predictions while being easily usable. 

\section{Methods}
In the following we provide a brief overview of PUUPL and direct the reader to \citet{puupl} for a comprehensive description.

PUUPL (Figure~\ref{fig:antiprocpath}b) is a PUL framework based on an iterative pseudo-labeling loop where the predictions of a baseline model on unlabeled samples, so-called pseudo-labels, are used as training targets to train a new model.
As this model is trained on a larger dataset with more labeled examples, it will perform better than the original model, as long as the pseudo-labels are correct.
PUUPL ensures this by choosing which samples to pseudo-label based on the predictive epistemic uncertainty of an ensemble of deep neural networks, since low uncertainty predictions are likely correct.
By iteratively repeating this loop of training a model and expanding the training set by pseudo-labeling some unlabeled examples it is possible to eventually improve the predictive performance.

One such scenario is when a dataset is highly imbalanced, i.e., it contains few labeled positive samples and most of the unlabeled samples belong to the negative class.
Cleavage prediction suffers from this issue, as the length of MHC-I-bound peptides suggests a cleavage probability of about 10\%.
In this situation, \citet{puupl} showed that PUUPL can greatly improve both predictive accuracy and calibration over PU risk estimators specialized for imbalanced data~\citep{imbnnpu} thus motivating our choice for this framework.

\section{Experimental protocol}

\begin{table*}[t]
\centering
\begin{tabular}{rrrrr}
\toprule
& \multicolumn{2}{c}{N-terminal} & \multicolumn{2}{c}{C-terminal} \\
&  \multicolumn{1}{c}{AUROC} & \multicolumn{1}{c}{AUPRC}
&  \multicolumn{1}{c}{AUROC} & \multicolumn{1}{c}{AUPRC} \\
\cmidrule(l){1-1}\cmidrule(l){2-3}\cmidrule(l){4-5}
NC 20S & 52.72  \err{0.02} & 18.85\err{0.02} & 66.07 \err{0.02} & 27.51\err{0.01}  \\
NC C term & 50.99 \err{0.02} & 18.68\err{0.02} & 81.53 \err{0.01} & 46.39\err{0.04} \\
NetCleave & 49.27 \err{0.02} & 17.54\err{0.02} & 79.61 \err{0.01} & 42.06\err{0.02} \\
\cmidrule(l){1-1}\cmidrule(l){2-3}\cmidrule(l){4-5}
imbnnPU & 75.15 \err{0.06} & 40.01\err{0.39} & 83.99 \err{0.06} & 57.37\err{0.33} \\
PUUPL & \textbf{78.00} \err{0.06} & \textbf{44.91\err{0.03}} & \textbf{87.20} \err{0.04} & \textbf{61.07\err{0.44}} \\
\bottomrule
\end{tabular}
\caption{\small 
Average standard error of area under the ROC curve (AUROC) and area under the precision-recall curve (AUPRC) on both datasets for three baselines (above), a PUL risk estimator and PUUPL.
}
\label{tbl:proc}
\end{table*}

\paragraph{Dataset}
We collected a dataset of 294,615 MHC-I epitopes from IEDB~\citep{iedb} and 89,853 from the Human MHC Ligand Atlas~\citep{hlala}.
To identify the potential progenitor protein of each epitope, we used BLAST~\citep{blast} and filtered for epitopes with an unique progenitor protein resulting in  258,424 data points.
Through the progenitor protein, we recovered the residues preceding the N-terminus and following the C-terminus of the epitope, thus providing context for the cleavage predictor.
We generated two separate datasets based exclusively on N- or C-termini cleavage sites, as it is known that the biological signal differs in these two situations~\citep{Schatz2008}.
We extracted ``decoy`` samples by considering cleavage sites located within three residues of the experimentally-determined terminus.
As discussed previously, it is unknown whether cleavage could or could not have happened at those positions, hence we treat such decoys as unlabeled in our PUL training procedure.
As they were all natural sequences, these decoys were unlikely to bias the models in any way.
The final datasets were then composed of 1,285,659 and 1,277,344 samples with 229,163 and 222,181 positives for the N- and C-terminus datasets respectively.

\paragraph{Model}
Each sample contains ten residues, six to the left and four to the right of the cleavage site.
The amino acids were one-hot encoded, resulting in a binary vector of 240 components for each cleavage site in the dataset.
We used an ensemble of multilayer perceptrons (MLPs) with batch normalization layers and relu activation.


\paragraph{Optimization}
We used Hyperband~\citep{Li2017HyperbandAN} with $\eta=3$ and $S=4$ to optimize PUUPL's hyperparameters, as well as the number and size of hidden layers and training regime of the MLPs on the C-terminus dataset.
Each hyperparameter combination was tested with ten-fold cross-validation, where a validation set of 50,000 samples was used for early stopping and a separate test set of the same size for the final scoring.

\paragraph{Evaluation}
As evaluation criterion we used the AUROC between positive and unlabeled samples (PU-AUROC), as previous work~\citep{Menon2015LearningFC,Jain2017RecoveringTC} has shown that higher PU-AUROC directly translates to higher AUROC on fully labeled data.
Note that, as we do not know true negatives, traditional metrics to evaluate classification performance such as accuracy, F1, precision, recall, etc. are not applicable in this case.
For imbnnPU and PUUPL we ran ten-folds cross-validation and used the statistical test proposed by~\citet{LeDell2015ComputationallyEC} to estimate the AUROC, its standard error and confidence intervals. 

As external baselines we consider NetChop~\citep{Nielsen_2005} and NetCleave~\citep{netcleave}, evaluating their predictions on ten random bootstraps of our dataset.
We also show evaluation scores for the imbnnPU loss~\citep{imbnnpu}, commonly used for PUL on imbalanced datasets.

\section{Results}
Both PUUPL and the imbalanced nnPU loss achieved lower performance on the N-terminals dataset, confirming previous observations that this predictive task is harder due to the biological processes involved~\citep{Schatz2008}.
On the C-terminal dataset, the imbalanced nnPU loss improved performance by 2.5 and 4.4 points compared to NetChop and NetCleave respectively, and PUUPL added a further 3.2 points reaching 87.2\% AUROC (Table~\ref{tbl:proc}).
In both datasets the difference in AUROC between imbnnPU and PUUPL was statistically significant at a significance of 1\%: the confidence intervals are $[74.99,75.32]$ and $[77.85,78.15]$ for N-terminals, and  $[83.85,84.14]$ and $[87.08,87.32]$ for C-terminals.
Note that both NetChop and NetCleave were only trained on C-terminals cleavage sites in the original publication, thus explaining their random predictions on the N-terminals dataset.
While the area under the precision-recall curve (AUPRC) is difficult to interpret in a PU setting, in an imbalanced scenario such as ours it can be used to ensure that the AUROC is not misleadingly overinflated.

\section{Conclusion}
We constructed a new dataset for proteasomal cleavage on both N- and C-terminals based on MHC-I ligands and trained an ensemble of MLP predictors with a pseudo-labeling framework for PUL~\citep{puupl}.

This improved performance by 5.6 points reaching 87\% AUROC on C-terminal cleavage sites and enabled novel cleavage predictions for N-terminals, which were not considered by the previous state-of-the-art.
Furthermore, our predictor only uses ten residues around the cleavage site, thus being more efficient in and requiring less data for generating predictions compared to NetChop.

In conclusion, accurate and efficient proteasomal cleavage predictors can be incorporated in epitope-based cancer vaccine design frameworks to improve vaccine efficacy at low doses, reducing deployment costs of personalized immunotherapies for cancer.

\acks{
E. D. was supported by the Helmholtz Association under the joint research school ”Munich School for Data Science - MUDS” (Award Number HIDSS-0006).
B. S. acknowledges financial support by the Postdoctoral Fellowship Program of the Helmholtz Zentrum M{\"u}unchen.
}

\bibliography{main}

\end{document}